\documentclass[a4paper,11pt]{article}
\usepackage{indentfirst}
\usepackage{graphicx,epsfig,url,xcolor,hyperref}
\usepackage{graphicx}
\usepackage{dcolumn}
\usepackage{epstopdf}
\usepackage{fixmath}
\usepackage{amsmath}
\usepackage{textcomp}
\usepackage{authblk}
\usepackage[a4paper, total={6in, 8in}]{geometry}
\usepackage{cite}
\usepackage[normalem]{ulem}
\usepackage{caption}
\usepackage{subcaption}

\title{Stability of Neutron Star and Cosmological Constant}
\author[1]{Naveen K. Singh}
\affil[1]{Sir P.T. Sarvajanik College of Science, Surat 395001, Gujarat, India}
\author[2]{Gopal Kashyap \thanks{\href{mailto:gplkumar87@gmail.com}{gplkumar87@gmail.com}}}
\affil[2]{Department
of Physics, School of Advanced Sciences, Vellore Institute of Technology, Vellore, Tamil Nadu 632014, India}
\date{}
\begin{document}

\vspace{10mm}

\maketitle

\begin{abstract}
We derive the equation for pressure within a neutron star, taking into account a non-zero cosmological constant ($\Lambda$). We then examine the stability of the neutron star’s equilibrium state in the presence of cosmological constant. Our analysis shows that the theorem used to assess the stability of stellar structures at equilibrium remains applicable to neutron stars even when a cosmological constant is considered. We further numerically solve the stellar structure equations and determine the mass of neutron star using different equations of state (EOS). Moreover, we observe that the value of the cosmological constant ($\Lambda \geq 10^{-11} \rm m^{-2}$) causes a significant change in the mass-radius relationship of neutron stars.
\end{abstract}

\section{Introduction}
Neutron stars can be observed in the entire electromagnetic spectrum, from the long wavelengths of radio waves to the short wavelengths of gamma rays \cite{Vidana:2018lqp}. Unlike other compact objects, neutron stars hold special significance due to their observability. They are characterized by extremely high density ($10^{13}-10^{15} \mbox{gm}/\mbox{cm}^{3}$), strong magnetic fields (around $10^{12} \mbox{Gauss}$), and small radii on the order of several kilometers. Neutron stars are formed after the supernova explosion of a progenitor star with a mass between 8 and 25 solar masses ($8-25  \mbox{M}_{\odot}$). Despite the intense gravitational pull within a neutron star, stability is achieved through neutron degeneracy pressure. Further research on neutron stars can be found in references\cite{Routaray:2024lni,Magnall:2024ffd,Parmar:2024qff}. \\
 
 The discovery of pulsars in 1967 surprised scientists, as they appeared in regular pulses. It was later confirmed that pulsars are rapidly rotating neutron stars. This rotation causes the magnetic field axis to move around the star's rotation axis, generating synchrotron radiation that appears as periodic pulses observable from Earth. The study of such neutron stars has since become essential for understanding high-energy astrophysical phenomena. Additionally, accurately estimating the matter density of the universe requires a clear understanding of the masses of various compact objects, including white dwarfs, neutron stars, and black holes, which contribute significantly to the universe’s overall mass. In this paper, we estimate the mass of neutron stars based on their equation of state. This calculation also requires solving the differential equation for pressure within neutron stars.\\

Cosmological data confirm that the universe is expanding at an accelerating rate \cite{Perlmutter,Riess}. This acceleration suggests the presence of dark energy, a form of energy with negative pressure. Various models have been proposed to explain dark energy, including modified gravity theories such as f(R) models \cite{Sotiriou:2008rp,Starobinsky:2007hu,Nojiri:2006gh}, scalar field theories, and the $\Lambda$-CDM (cold dark matter) model \cite{Chiba:2005tj,Tsujikawa:2006mw,Tsujikawa:2005ju,Copeland:2006wr}. Among these, cosmological data strongly support the $\Lambda$-CDM model, where the cosmological constant $\Lambda$ serves as a source of dark energy. The energy density associated with the cosmological constant $\rho_\Lambda$ (contributing approximately 70\%) and cold dark matter (about 25\%) together provide a compelling explanation for current cosmological observations \cite{Planck:2018vyg}.

In this paper, we incorporate the cosmological constant into the theoretical framework of neutron stars. Although stability analyses considering the cosmological constant have been conducted in previous studies \cite{Bordbar:2015wva,Afifah:2020vnp,R:2024ovl},  we revisit this analysis using the stability theorem presented in Ref. \cite{Weinberg:1972kfs}. Additionally, we estimate the mass of neutron stars while accounting for the effects of the cosmological constant, providing insights into how this factor influences neutron star structure and stability.
 
This paper is structured as follows: Section 2 details the derivation of the stellar structure equations, incorporating a non-zero cosmological constant. In Section 3, we examine the stability theorem in the presence of the cosmological constant. Section 4 discusses the solution of the stellar equations for a given equation of state (EOS) without the cosmological constant. Section 5 presents the effects of the cosmological constant on stellar structure. Finally, Section 6 summarizes our findings and discusses the implications for understanding the potential role of the cosmological constant in neutron star physics.
 
 \section{Stellar Structure Equations with Cosmological Constant}
 In this section, we begin with the Einstein equations  that describe neutron stars, including the cosmological constant. These equations can be expressed as follows:
\begin{eqnarray}
 R_{\mu \nu} -\Lambda g_{\mu \nu} = -8 \pi G \left(T_{\mu \nu} 
  - \frac{1}{2} g_{\mu \nu} T^{\lambda}_{\lambda} \right),
\end{eqnarray}
where $R_{\mu\nu}$, $\Lambda$ and $T_{\mu \nu}$ are Ricci tensor, cosmological constant, and energy-momentum tensor of neutron star matter, respectively. We define the proper pressure and proper energy density of ordinary matter fluid inside the star as p and $\rho$, respectively. The velocity four vector $U^{\mu}$ satisfies the relation $g^{\mu \nu} U_{\mu} U_{\nu}=-1$. The energy-momentum tensor $T_{\mu \nu}$ is defined as   $T_{\mu \nu} = p g_{\mu \nu} + (p+ \rho)U_{\mu} U_{\nu}$. We consider the fluid to be at rest so that $U_r = U_\theta = U_\phi =0$ and $U_t = - (-g^{tt})^{-1/2}$ for spherically symmetric static star. For such a compact star, we consider the following metric,
\begin{eqnarray}
 g_{tt} = - B(r), \ \ g_{rr} = A(r), \ \ g_{\theta \theta} = r^2, \ \ g_{\phi \phi} = r^2 \sin^2 \theta  \ \ 
\end{eqnarray}
with $g_{\mu\nu}=0$ for $\mu \neq \nu$. Using this metric, t-t, r-r, and $\theta$-$\theta$ components of Einstein equation can be written as,

\begin{eqnarray}
 -\frac{B''}{2 A} + \frac{B'}{4 A} \left(\frac{A'}{A}+ \frac{B'}{B}\right) - \frac{B'}{r A} = - 4 \pi G B\left(\rho+ \frac{\Lambda}{4 \pi G} + 3 p \right), \label{EEq1}\\
 \frac{B''}{2 B} - \frac{B'}{4 B}\left(\frac{A'}{A}+ \frac{B'}{B}\right) - \frac{A'}{r A} = - 4 \pi G A \left(\rho -\frac{\Lambda}{4 \pi G} - p\right), \label{EEq2}\\
 -1 + \frac{1}{A} + \frac{r}{2 A} \left( \frac{B'}{B}-\frac{A'}{A}\right) = - 4 \pi G r^2 \left( \rho -\frac{\Lambda}{4 \pi G} -p \right)\label{EEq3}.   
\end{eqnarray}
From the conservation of Energy-momentum tensor, i.e., $T^{\mu\nu}_{\ \ ;\nu} =0$, we have,
\begin{eqnarray}
 \frac{B'}{B} = - \frac{2 p'}{\rho + p}  \label{EqnB}.
\end{eqnarray}
Eliminating B from Eqs (\ref{EEq1}), (\ref{EEq2}) and (\ref{EEq3}), we obtain
\begin{eqnarray}
 \left(\frac{r}{A}\right)' = 1 - 8 \pi G r^2 \left(\rho - \frac{\Lambda}{8 \pi G}\right).
\end{eqnarray}
Solution for A can be written as
\begin{eqnarray}
 A(r) = \left(1 -  \frac{2 G M(r)}{r}\right)^{-1} \label{solA},
\end{eqnarray}
where $M(r)$ is defined as,
\begin{eqnarray}
 M(r) = \int^{r}_{0} 4 \pi r'^2 \left(\rho(r') - \frac{\Lambda}{8 \pi G} \right) dr'. \label{EqnM}
\end{eqnarray}
Alternatively, 
\begin{eqnarray}
 \frac{d M}{dr} = 4 \pi r^2 \left(\rho(r) - \frac{\Lambda}{8 \pi G}\right). \label{diffEqM}
\end{eqnarray}
Now, using Eqs. \ref{EqnB}, \ref{solA} and \ref{EEq3}, we obtain,
\begin{eqnarray}
  \frac{p'}{(\rho + p)} = -\frac{ G M(r)}{r^2} \left(1- \frac{2 G M(r)}{r}\right)^{-1} \left(1 + \frac{ \Lambda r^3}{2 G M(r)}  + \frac{4 \pi r^3 p}{M(r)}\right) \label{diffEq}
\end{eqnarray}
From Eq. \ref{EqnB} and \ref{diffEq}, we obtain the solution of B as 
\begin{eqnarray}
 B = \exp \left(- \int^C_r \frac{2 G}{r'^2} \left(M(r')+ \frac{\Lambda r'^3}{2 G} + 4 \pi r'^3 p(r')\right)\left(1 - \frac{2 G M(r')}{r'}\right)^{-1} dr'\right) . \label{SolB}
\end{eqnarray}
Here, $C$ is a constant. Outside the star $M(r)$ becomes,
\begin{eqnarray}
 M(r) &=& \int^R_0 4 \pi r'^2 \rho(r') d r' - \frac{\Lambda}{2 G} \int^{r}_0 r'^2 dr', \\
 &=& M(R) -   \frac{\Lambda}{2 G} \int^{r}_0 r'^2 dr'
\end{eqnarray}
where, $\rho(r)=0$ for $r>R$ with $R$ is radius of star. $M(R)$ is defined as mass of star. Outside the star, pressure is zero and hence solution for $B$ can be written as
\begin{eqnarray}
 B &=& \exp \left(-\int^C_r \frac{\left(\frac{2 G M(R)}{r'^2} +\frac{2 \Lambda r'}{3}\right)}{\left(1- \frac{2 G M(R)}{r'} + \frac{ \Lambda r'^2}{3}\right)}\right)\\
 &=& \exp \left(- \left(\ln \left(1 - \frac{2 G M(R)}{r'}+ \frac{\Lambda r'^2}{3} \right)_{r'=C} - \ln \left(1 - \frac{2 G M(R)}{r'}+ \frac{\Lambda r'^2}{3} \right)_{r'=r}\right)\right) \nonumber \\
\end{eqnarray}
We choose constant $C$ such that
\begin{eqnarray}
  \left(1 - \frac{2 G M(R)}{r'}+ \frac{\Lambda r'^2}{3} \right)_{r'=C} = 1.
\end{eqnarray}
Thus, the solution of $B(r)$ outside the star, takes the form as
\begin{eqnarray}
 B(r) =  1 - \frac{2 G M(R)}{r}+ \frac{\Lambda r^2}{3},
\end{eqnarray}
and solution of $A(r)$ takes its form as,
\begin{eqnarray}
 A(r) = \left( 1 - \frac{2 G M(R)}{r}+ \frac{\Lambda r^2}{3} \right)^{-1},
\end{eqnarray}
leading to $A(r) B(r) =1$.
\section{Stability Test in the Presence of Cosmological Constant} 
In the absence of cosmological constant, a  theorem to test the stability of solution is applicable to a star.  The theorem is as follows \cite{Weinberg:1972kfs}: 
A particular stellar configuration, with uniform entropy per nucleon and chemical composition, will satisfy the differential equation of pressure for equilibrium if and only if the quantity $M(r) = \int 4 \pi r^2 \rho(r) dr$ is stationary with respect to all variations of $\rho(r)$ that leave unchanged the quantity,
\begin{eqnarray}
 N = \int 4 \pi r^2 n(r) \left(1 - \frac{2 G M(r)}{r}\right)^{-1/2} dr,
\end{eqnarray}
where, $N$ is the number of nucleons in the star. The Lagrange multiplier method provides a relation $\delta M - \lambda \delta N = 0$ for a given variation $\delta \rho(r)$. This theorem is proven in Ref. \cite{Weinberg:1972kfs}. We will test the theorem in the presence of cosmological constant. \\

In the presence of cosmological constant, from Eq. (\ref{EqnM}), the variation of $\rho(r)$ leads to
\begin{eqnarray}
 \delta M = \int^\infty_0 4 \pi r^2 \delta \rho(r) d r
\end{eqnarray}
Using Lagrange multiplier, we obtain:
\begin{eqnarray}
 \delta M - \lambda \delta N =  \int^\infty_0 4 \pi r^2 \Bigg[ 1 - \frac{\lambda n(r)}{p(r)+ \rho(r)} \left(1 - \frac{2 G M(r)}{r}\right)^{-1/2} \nonumber \\
 -\lambda G \int^\infty_r 4 \pi r' n(r') \left(1- \frac{2 G M(r')}{r'}\right)^{-3/2} dr'\Bigg] \delta \rho(r)dr, \label{deltaMN}
\end{eqnarray}
where we have used the formula to interchange the integral and we also used the fact that entropy per nucleon  is fixed under the variation of energy density, i.e.,
\begin{eqnarray}
 \delta \left(\rho/n\right) + p \delta (1/n)=0
\end{eqnarray}
which leads to,
 \begin{eqnarray}
 \delta n(r) =\frac{n(r)}{p(r)+ \rho(r)} \delta\rho(r).
\end{eqnarray}
Also because of uniform entropy per nucleon, we have,
\begin{eqnarray}
 \frac{d}{dr}\left(\frac{\rho}{n}\right) + p\frac{d}{dr}\left(\frac{1}{n}\right) =0,
\end{eqnarray}
which can be written as,
\begin{eqnarray}
 n'(r) = \frac{n(r) \rho'(r)}{p(r)+\rho(r)}.
\end{eqnarray}

$M - \lambda N$ can be  stationary with respect to variation of $\rho(r)$, if $\delta M - \lambda \delta N =0$, i.e.,
\begin{eqnarray}
 \frac{1}{\lambda} = \frac{n(r)}{p(r)+\rho(r)} \left(1 - \frac{2 G M(r)}{r}\right)^{-1/2} + G \int^\infty_r 4 \pi r' n(r') \left(1- \frac{2 G M(r')}{r'}\right)^{-3/2} dr'  \label{lambdainv}
\end{eqnarray}
Differentiating Eq. \ref{lambdainv} with respect r, we obtain,

\begin{eqnarray}
 \frac{n'}{p + \rho} - \frac{n (p'+\rho')}{(\rho + p)^2} + \Bigg[\frac{G n}{p+\rho} \left( 4 \pi r \rho-\frac{\Lambda r}{2 G} - \frac{M(r)}{r^2}\right) - 4 \pi G r n\Bigg] \left(1- \frac{2 G M(r)}{r}\right)^{-1} =0 \nonumber \\ 
\end{eqnarray}
Using $n' = \frac{\rho' n}{\rho + p}$ and simplifying we obtain,
\begin{eqnarray}
 \frac{p'}{\rho + p} = - \left(\frac{\Lambda r}{2} + \frac{G M(r)}{r^2} + 4 \pi G r p\right) \left(1 -  \frac{2 G M(r)}{r}\right)^{-1}
\end{eqnarray}
which is same as Eq. \ref{diffEq}. Therefore, the theorem still holds in the presence of cosmological constant. The stability of this equilibrium state will be determined by the sign of the second-order variation $\delta M$ with respect to $\delta \rho$; it will be stable if $\delta M$ is positive and unstable if $\delta M$ is negative.

\section{Mass-Radius Relation of Neutron Star}
In this section, we discuss the numerical solution of differential equation Eq.~\ref{diffEq} with $\Lambda=0$, known as the Tolman-Oppenheimer-Volkoff (TOV) equation. Solving TOV equation provides key stellar observables, including the mass and radius of the neutron star. For a given central mass density, numerically integrating the TOV equation until the pressure reaches zero defines the star’s radius. The total mass of the star is then determined by integrating the mass density up to this radius.

To solve the TOV equation, it is essential to know the equation of state (EOS), which provides the relationship between pressure and energy density for a given star. The EOS is determined by the nature of the star’s energy density components. Quantum degeneracy, a key effect in compact stars, plays an important role in establishing degeneracy pressure in white dwarfs and neutron stars. The theoretical framework of fermion degeneracy pressure provides detailed insights into the EOS. In a white dwarf, the degeneracy pressure arises from degenerate electrons, while in a neutron star, it results from degenerate neutrons. Consequently, the calculations for degeneracy pressure in both types of stars are similar, differing only due to the masses of electrons and neutrons. Once the EOS of the neutron star is established, we can solve the TOV equation to find the mass and radius of stars.

 One important aspect of astrophysics that provides insights into the internal structure and the equation of state of compact star is the mass-radius curve. By analyzing this relationship, we can better understand the physics governing neutron stars and constrain models that describe their formation and evolution. To create these curves, it is essential to solve the TOV equations for various central densities. Consideration of different components of matter leads to different EOS and hence leads to different radii and masses of the star.

Considering an ideal Fermi gas of neutrons, for $\rho(0) \ll \rho_c$, where $\rho(0)$ is the central density and $\rho_c = 6.11 \times 10^{15} \, \text{gm}/\text{cm}^3$ is the critical density, the mass of the neutron star is given by $ M = 2.7 \, \left( \frac{\rho(0)}{\rho_c} \right)^{1/2} M_{\odot} $, and the radius is $ R = 3.653 \, \left( \frac{\rho(0)}{\rho_c} \right)^{1/6} R_0 $, where $ R_0 = 3.0 \, \text{km} $. For $\rho(0) \gg \rho_c$, as $\rho(0) \to \infty$, the asymptotic values are $ M_{\infty} = 0.341 \, M_{\odot} $ and $ R_{\infty} = 1.06 \, R_0 $ \cite{Weinberg:1972kfs}.

If we consider a neutron star as a polytrope, the mass density (or matter energy density) $\rho(r)$ is related to the pressure $p(r)$ by the relation $ p(r) = K \rho^\gamma $ \cite{Weinberg:1972kfs}, where $ K $ is a dimensional constant and $\gamma$ is the polytropic index. Redefining the variables as $ M = M_\odot m $, $ r = R_\odot \xi $, and $ \rho = \rho(0) \theta $, where $ M_\odot $, $ R_\odot $, and $ \rho(0) $ represent the solar mass, solar radius, and central matter density of the star, respectively. In term of these variables, the TOV equation, assuming $\Lambda = 0$, can be written as,
\begin{eqnarray}
\frac{d \theta}{d \xi} &=& - C_1 \frac{m \theta^{2 - \gamma}}{\xi^2} \left( 1 + C_2 \theta^{\gamma - 1}\right) \left( 1 + C_3 \frac{\theta^\gamma \xi^3}{m}\right)\left(1 - C_4 \frac{m}{\xi}\right)^{-1},
\end{eqnarray}
\begin{eqnarray}
\frac{d m}{d \xi} &=& C_5 \xi^2 \theta,
\end{eqnarray}
where the constants are defined as,
\begin{eqnarray}
 C_1 = \frac{C_4}{2 \gamma C_2}, \ \  C_2 = K \rho(0)^{\gamma -1} = p(0)/\rho(0), \ \  C_3 = C_5 C_2, \ \
 C_4 = \frac{2 G M_\odot}{R_\odot}, \ \ 
 C_5 = \frac{4 \pi R_\odot^3 \rho(0)}{M_\odot}. \nonumber \\ 
\end{eqnarray}

If the polytropic index of the star is known, we can solve these equations to determine the star's mass and radius. In this paper, we directly utilized the online service CompOSE\footnote{https://compose.obspm.fr} \cite{compose}, which provides data tables for various state-of-the-art equations of state. This resource allows us to access EOS models based on updated nuclear physics data, enabling precise calculations of neutron star properties.

\begin{figure}[ht]
         \centering
         \includegraphics[width=\textwidth]{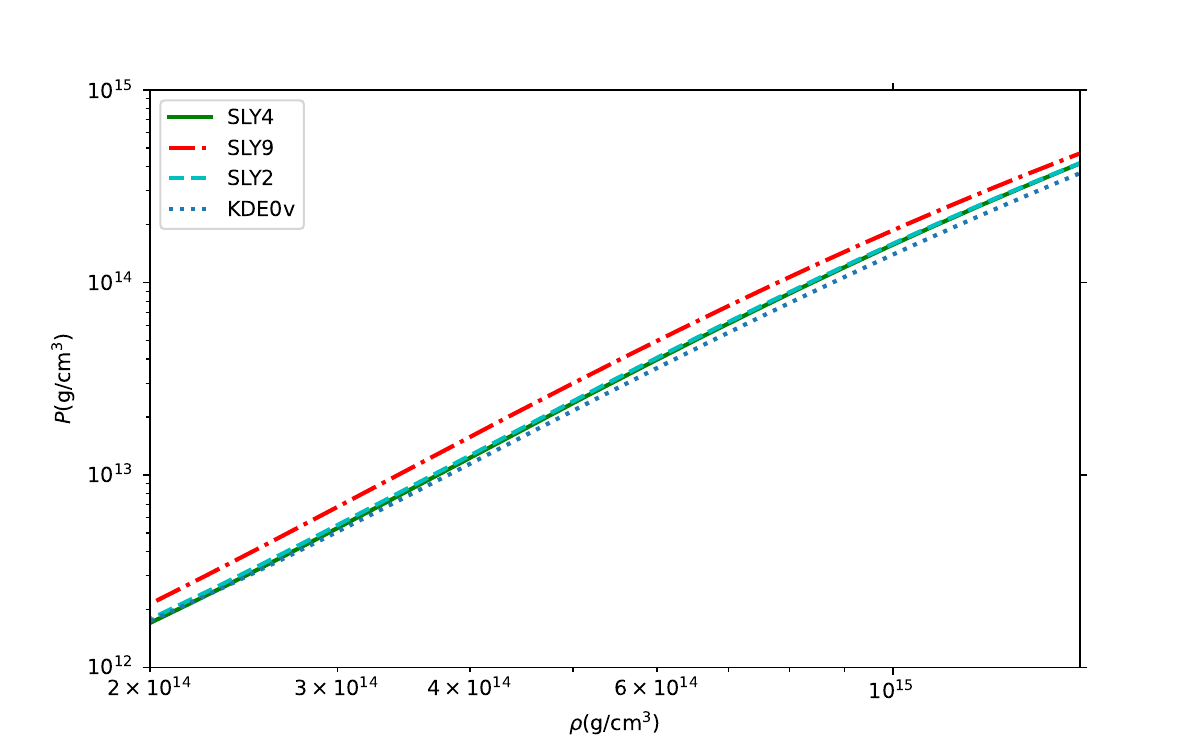}
     \caption{Equation of state for nuclear matter used in this work.}
     \label{Eos}
\end{figure}

\begin{figure}[ht]
     \begin{subfigure}[b]{0.5\textwidth}
         \centering
         \includegraphics[width=8cm]{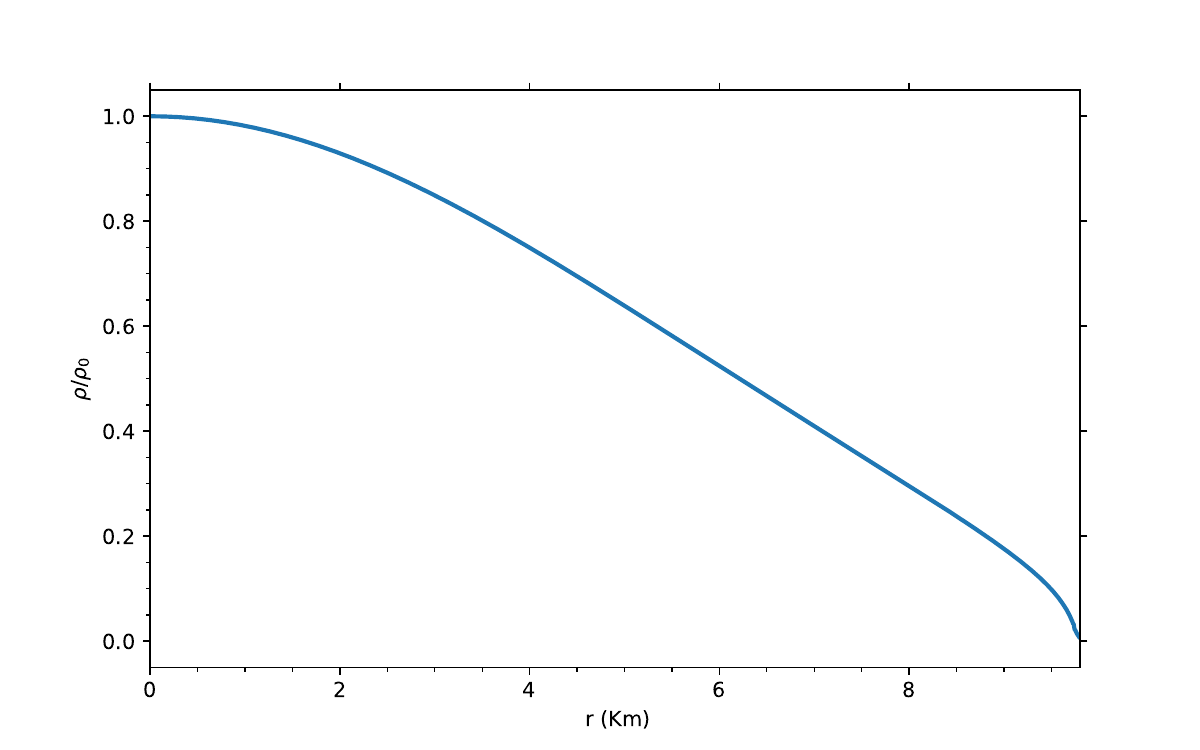}
     \end{subfigure}
     \begin{subfigure}[b]{0.5\textwidth}
         \centering
         \includegraphics[width=8cm]{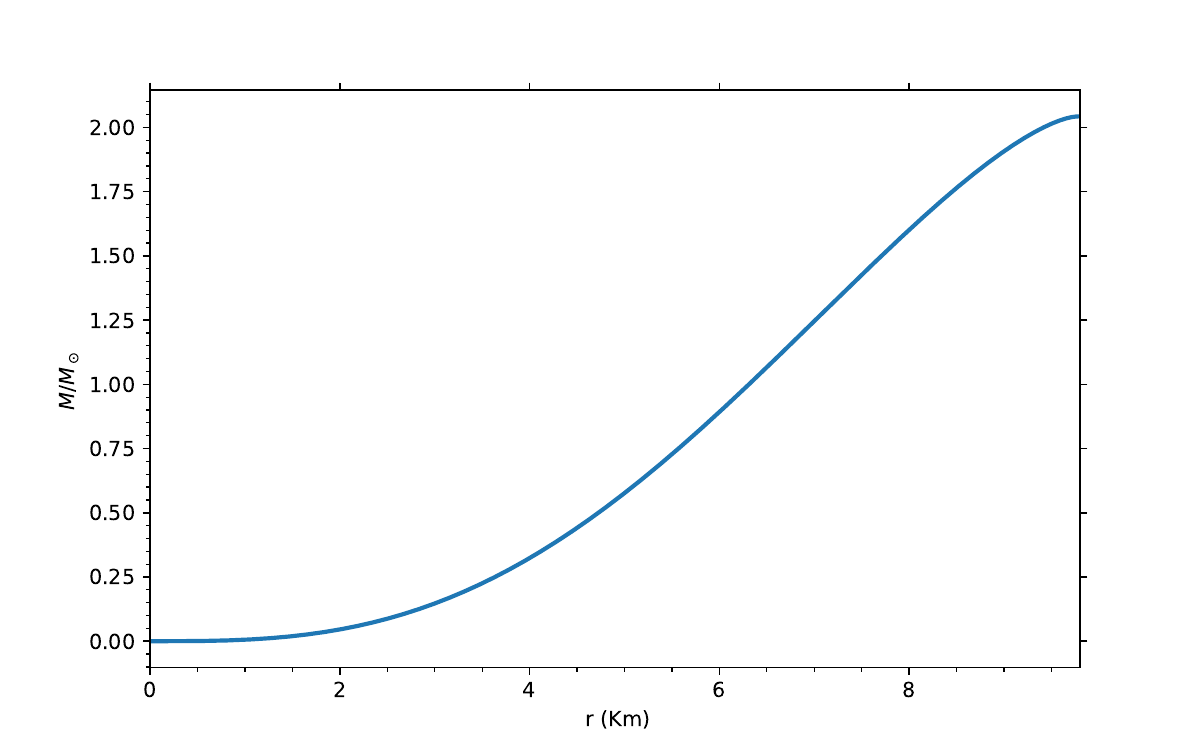}
     \end{subfigure}
      \caption{Plot of mass density and mass of neutron star with respect to radial distance from the centre with $\Lambda=0$ for SLY4 EOS. }
      \label{density_mass}
\end{figure}

For this study, we selected four different models for the equation of state of nuclear matter: SLY4, SLY9, SLY2, and KDE0v, all of which incorporate the Skyrme interaction \cite{GR_2015,DLNP_2009, Araujo:2024txe,Dutra:2012mb}. These EOSs are shown in Fig. \ref{Eos}. 
In these models, the effective mass and nuclear potential are key parameters used to characterize the bulk properties of nuclear matter. Using the SLY4 EOS, we solved the TOV equation to examine how the mass density, $\rho(r)$, and neutron star mass, $m(r)$, vary with radial distance for a fixed central mass density. The results are shown in Fig.~(\ref{density_mass}).

\begin{figure}[ht]
         \centering
         \includegraphics[width=\textwidth]{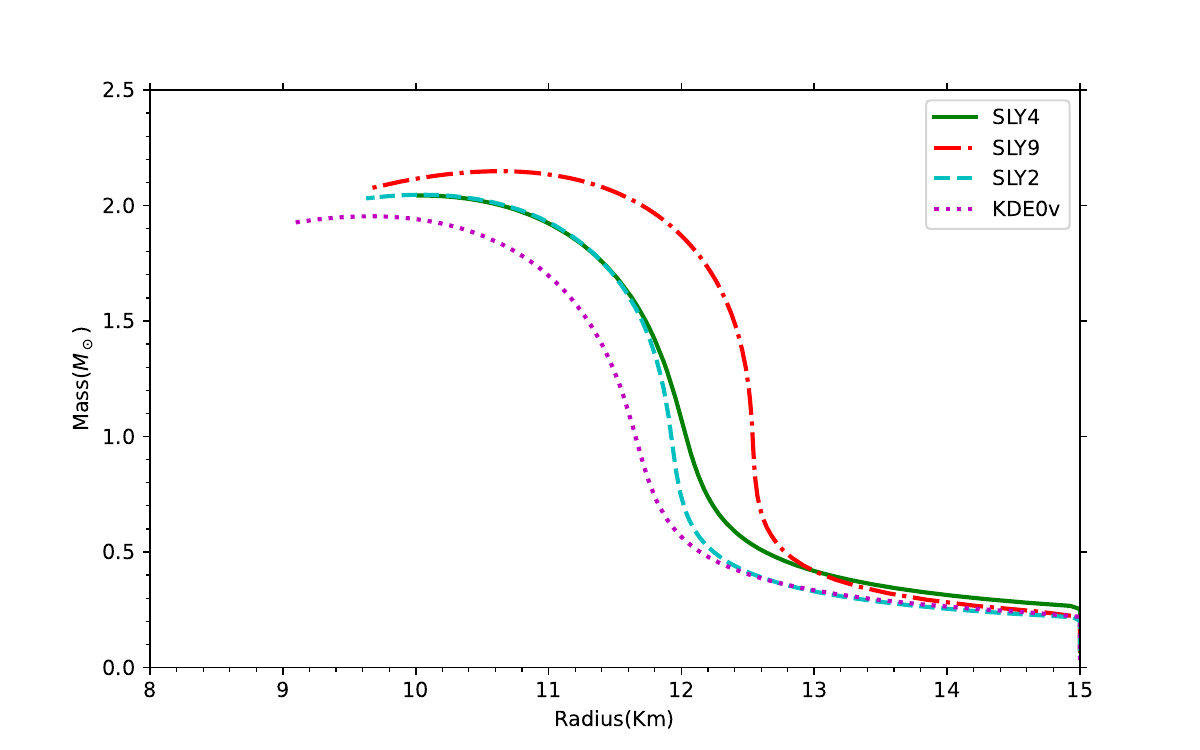}
      \caption{ Mass-radius relationship for the selected EOSs with $\Lambda=0$.}
      \label{M-R}
\end{figure}

In Fig.~\ref{M-R}, we showed the mass-radius relationship for neutron stars, derived from the equations of state presented in Fig.~\ref{Eos}. This relationship is calculated without including any cosmological constant, focusing solely on the effects of the EOS on the neutron star's structural properties. Each EOS represents different assumptions about the interactions and compressibility of matter at extreme densities, influencing the resulting star's mass and radius. By comparing these mass-radius curves, we can observe how variations in the EOS affect the maximum stable mass and corresponding radius of neutron stars, offering insights into the dense matter properties within the star.

\section{The effect of cosmological constant on stellar structure}
\begin{figure}[ht]
         \centering
         \includegraphics[width=\textwidth]{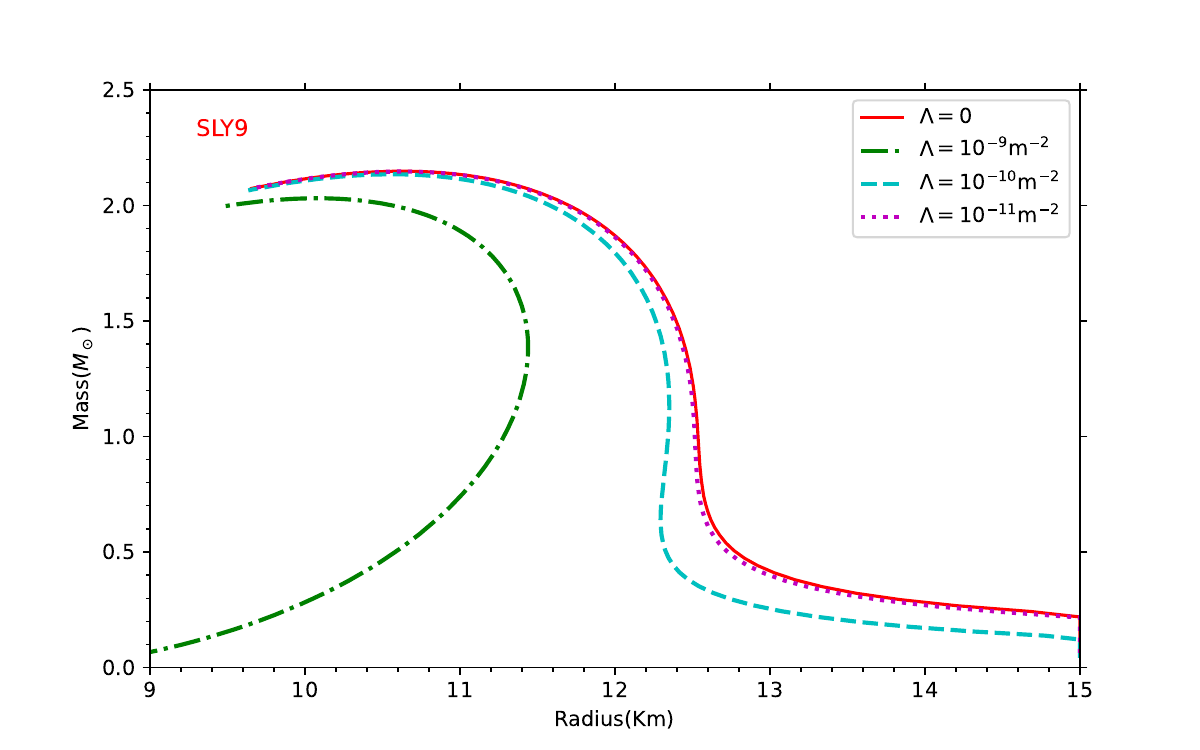}
     \caption{Mass-Radius relation for SLY9 EOS with the different values of cosmological constant.}
     \label{M-R_lambda}
\end{figure}

A non-zero cosmological constant contributes a small, repulsive term to gravity, which can affect the star’s equilibrium configuration. For sufficiently large values of $\Lambda$, this repulsive effect can cause notable shifts in the mass-radius relationship. These changes provide insight into how dark energy might impact astrophysical objects on smaller scales, connecting cosmological effects to stellar structure. To examine the impact of the cosmological constant on stellar structure, we used the SLY9 EOS model and calculated the mass-radius relationship for various values of $\Lambda$. The results are shown in Fig.~\ref{M-R_lambda}.

As shown in Fig.~\ref{M-R_lambda}, increasing $\Lambda$ causes a notable shift in the mass-radius relationship toward lower radii and masses. This indicates that a non-zero cosmological constant not only reduces the star’s size but also impacts its maximum stable mass. This occurs because the vacuum energy associated with $\Lambda$ significantly affects the equation of state across the entire density range.  Such dependencies highlight the role of dark energy in compact astrophysical objects, implying that even a small cosmological constant could affect neutron star properties in measurable ways. Understanding these effects may help in refining the models for neutron star interiors and offer new constraints on the values of $\Lambda$ from stellar observations.
\section{Conclusions}
In this work, we extended the stability theorem for equilibrium configurations in neutron stars, accounting for a non-zero cosmological constant ($\Lambda$) while assuming uniform entropy per nucleon and chemical composition. Our results confirm that the stability theorem remains valid in the presence of a cosmological constant. We then analyzed the mass-radius relationship for neutron stars under various equation of state models, both with and without $\Lambda$. For the current observed value of $\Lambda$, there is no significant impact on the mass-radius relation. However, noticeable changes emerge when $\Lambda$ is set to values around $10^{-11} \, \text{m}^{-2}$, far larger than the observed cosmological constant.

For EOS models such as SLY2, SLY4, and SLY9, we observed that the maximum neutron star mass approaches approximately 2.0 $M_{\odot}$ with a radius of about 10 km, while the KDE0v model yields a maximum mass slightly below 2.0 $M_{\odot}$ and a radius under 10 km. With $\Lambda$ values ranging from $10^{-11} \, \text{m}^{-2}$ to $10^{-9} \, \text{m}^{-2}$, we observed significant variation in the maximum radius and maximum mass of neutron stars. These findings show the potential effect of large cosmological constant values on neutron star properties while affirming the stability framework for compact stellar objects. This study may refine models of neutron star interiors and contribute to constraints on the cosmological constant derived from observations of stellar properties.
\section*{Acknowledgements}
GK acknowledges the Vellore Institute of Technology for providing financial support through its Seed Grant (No.SG20230035), year 2023.

\bibliographystyle{JHEP}
 \bibliography{NeutronStar}

\end{document}